\documentclass[twocolumn,showpacs,prb]{revtex4}
\usepackage{graphicx}
\usepackage{amsmath}
\usepackage{amssymb}

\begin{document}

\title{Ferromagnetic redshift of the optical gap in GdN}
\author{H. J. Trodahl} \email{Joe.Trodahl@vuw.ac.nz}
\author{A. R. H. Preston}
\author{J. Zhong}
\author{B. J. Ruck}
  \affiliation{The MacDiarmid Institute for Advanced Materials and Nanotechnology, 
    School of Chemical and Physical Sciences, 
    Victoria University, 
    PO Box 600, Wellington 6140, New Zealand}
\author{N. M. Strickland}
  \affiliation{Industrial Research Ltd., 
    Lower Hutt, 
    PO Box 31310, Lower Hutt 5040, New Zealand}
\author{C. Mitra}
\author{W. R. L. Lambrecht}
  \affiliation{Department of Physics, Case Western Reserve University, 
    Cleveland, Ohio 44106-7079, USA}

\begin{abstract}
We report measurements of the optical gap in a GdN film at
temperatures from 300 to 6~K, covering both the paramagnetic and
ferromagnetic phases. The gap is 1.31~eV in the paramagnetic phase and
red-shifts to 0.9~eV in the spin-split bands below the Curie
temperature. The paramagnetic gap is larger than was suggested by very
early experiments, and has permitted us to refine a
(LSDA+\emph{U})-computed band structure. The band structure was computed 
in the full translation symmetry of the ferromagnetic ground state, 
assigning the paramagnetic-state gap as the average of the 
majority- and minority-spin gaps in the ferromagnetic state. 
That procedure has been further tested by a band structure in a 32-atom 
supercell with randomly-oriented spins. After fitting only the 
paramagnetic gap the refined band structure then reproduces our measured 
gaps in both phases by direct transitions at the X point.
\end{abstract}


\date{\today}

\maketitle

\section{Introduction}
The rare earth nitrides (RE-N) have recently attracted attention
following theoretical advances that have yielded credible band
structures in these strongly correlated
materials.\cite{lambrecht2000eso,ghosh2005ema,johannes2005mcb,larson2006esg,bhattacharjee2006ioc,larson2007esr,aerts2004hmi,duan2005sih,antonov2007xrm,chantis2007gwm2} However, despite
their simple rock-salt structure and their strongly localized
4\emph{f} states there are disagreements among various theoretical
treatments regarding the nature of their band structures in either the
ambient-temperature paramagnetic state or their magnetic ground
states. Interestingly, among the predictions one finds that some may
be half metals\cite{larson2007esr,aerts2004hmi,duan2005sih} which are of interest for spintronics
applications, though their magnetic order is limited to temperatures
below 70~K. However, even at ambient temperature the lattice constant 
varies systematically across the series, leading in turn to a systematically 
varying band structure and band gap. Thus the RE-N compounds may prove useful 
in a range of electronic and electro-optic applications.

The ionicity of these materials is manifest in their band structures,
so that their valence bands are of N 2\emph{p} and the conduction
bands RE 5\emph{d}, 6\emph{s} character. In the presence of partially
filled RE 4\emph{f} levels the exchange interaction
shifts the spin-split conduction and valence band edges in the
opposite sense, reducing the majority spin gap while the minority
spin gap opens.\cite{larson2007esr} Clearly a sufficiently large shift will
reduce the majority spin gap to zero, resulting in a half
metal. Besides the strong correlation effects affecting the 4\emph{f}
states, the gap is also affected by the usual underestimate of the gap
by the local density approximation (LDA). The latter is primarily due
to long-range Coulomb contribution which is over-screened by the
electron-gas screening used in LDA. In the present case, the
conduction band minimum consists of RE-$d_{t2g}$ states and the
underestimated gap can be corrected by shifting the RE-\emph{d} states
upward within the LSDA+\emph{U} method by introducing a $U_{d}$ even
though the underlying physics is rather different from the Hubbard
$U_{f}$ shifts.\cite{lambrecht2000eso,larson2007esr} The calculations require experimental
input concerning the gap in order to fix a value for the empirical
parameter $U_{d}$. In this paper we present transmission data that establish the
optical gaps of GdN in both magnetic states, which then allow us to
adjust the computed band structure.

The experimental description of RE-Ns is far from clear. Although
their NaCl structure is well established,\cite{hullinger1978hot} there is
remarkably little consensus in the literature concerning their
physical properties. Most are known to be magnetically ordered at low
temperature,\cite{vogt1993hot} though there remains much uncertainty about
transition temperatures, saturation moments and even in some cases
whether there is magnetic order at all and whether it is ferro- or
antiferro-magnetic.\cite{vogt1993hot,wachter1980mia,gambino1970mpa,li1997sgb,preston2007bso} There is also debate
about their temperature-dependent conductivity, with specific samples
of even nominally the same composition claimed variously as a metal,
semimetal or semiconductor.\cite{wachter1980mia,wachter1998bsp,degiorgi1990esy,dismukes1970,granville2006sgs,leuenberger2005gtf} Nitrogen
vacancies are common, and even in a material that is fundamentally a
semiconductor they may dope the material to degenerate carrier
densities.\cite{dismukes1970,granville2006sgs}

\begin{figure}
\centering{ 
  \includegraphics[width=8cm]{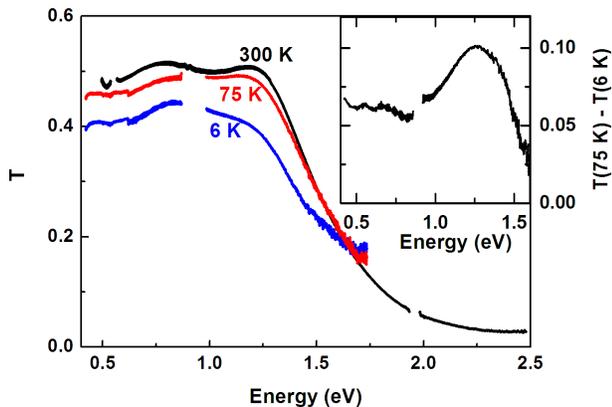} 
}
\caption{ (color online) Transmission through a 200~nm GdN/$\sim$200
  nm GaN film deposited on sapphire. The 300~K data were collected
  with the sample in vacuum, and for the 6 and 75~K data there were an
  additional four polythene windows in the path which sacrificed
  intensity above 1.5~eV and across a gap near 0.9~eV. The edge at
  1.3~eV is clearly visible in 75 and 300~K data, as is the tail of
  reduced intensity below that threshold at 6~K. The inset shows the
  transmission reduction in the ferromagnetic phase, with a clear
  indication of an inter-band edge at a lower energy than in the
  paramagnetic phase.}
\label{fig1}
\end{figure}

Optical transmission experiments have the potential to settle these
questions by providing a measurement of the optical (i.e., minimum
\emph{direct}) band gap, but here also the few reported measurements
have provided ambiguous results. Very early studies of the entire
range of rare earth nitrides (excepting the radioactive Pm) were based
on powder samples, which dictated that only diffuse reflection
measurements could be made.\cite{dismukes1970,busch1969poc2} Those data show only a weak
dip in their diffuse absorption ($A_{\mathrm{\textit{diff}}} =
1-R_{\mathrm{\textit{diff}}}$), with enhanced absorption at energies
both larger and smaller than the quoted gap, which was interpreted as
resulting from competition between inter-band and free-carrier
contributions to the optical conductivity. The gaps, all near 1~eV,
were taken as the weak minima between the two competing contributions
to the diffuse absorption, rather than by observations of any clear
absorption edges. In contrast gaps of 2 to over 4~eV\cite{sclar1964por,shalaan2006sao} are
likely to be associated with severe oxidation; Gd$_{2}$O$_{3}$ has a
5.2~eV gap.\cite{singh2004soa} We find that unprotected films become
transparent insulators within seconds after exposure to air.\cite{granville2006sgs}
The propensity of the RE nitrides to oxidize joins the presence of N
vacancies as the phenomena most responsible for ambiguous historical
results.

GdN, with its half-filled 4\emph{f} shell, is by far the most
thoroughly investigated of the RE-Ns. It has the highest Curie
temperature among the RE-Ns,\cite{vogt1993hot,granville2006sgs,schumacher1965mco,cutler1975sam,li1994mpo,li1997mpo,leuenberger2005gtf,leuenberger2006xrm,khazen2006frs}
with reports as high as 90~K.\cite{cutler1975sam} We and others have recently
reported a resistivity showing a magnitude and temperature dependence
that is characteristic of a semiconductor above T$_C$, followed by a
rapid fall as the temperature is lowered into the ferromagnetic
phase.\cite{granville2006sgs,leuenberger2005gtf} However, below 25K the 
resistivity again begins to increase in a form consistent with a 
reduced-gap semiconducting ferromagnetic state.\cite{granville2006sgs}

We recently reported visible-near IR measurements on GdN demonstrating
a clear ambient-temperature absorption edge near 1~eV, but without
sufficient data below 1~eV to make a reliable estimate of the
gap.\cite{granville2006sgs} In this paper we extend transmission measurements to
lower energy, which has permitted an unambiguous determination of the
optical gap. More importantly the measurements have been completed at
temperatures from ambient down to 6~K and signal clear band structure
changes at the Curie temperature.

\section{Experimental Details}
Thin films of GdN were grown, as described in more detail
earlier,\cite{preston2007bso,granville2006sgs,mckenzie2005tem} by deposition of Gd from an electron beam
heated source in the presence of $10^{-4}$~mbar of pure nitrogen
gas. The 200 nm-thick film was capped with $\sim$200~nm GaN, which we
have shown to be an effective barrier to reaction with the
atmosphere.\cite{granville2006sgs} Note that in GaN the inter-band edge is
3.4~eV,~\cite{koo2006pin} and there is only a small, approximately 
energy-independent absorption in the energy range probed in this study.

The GdN in the present study contains crystallites with
diameters of about 8 nm and has a resistivity of 0.15~$\Omega$~cm at
300~K, rising through a peak of 0.25~$\Omega$~cm at a Curie
temperature of 66~K and thereafter falling to a minimum of
0.18~$\Omega$~cm.  A carrier density of $10^{18}$~cm$^{-3}$ at 300~K
has been estimated by noting that the mean-free-path cannot be larger
than the radius of the crystallites.

Spectral measurements in the range of 0.2-2~eV were performed with a
Bomem model DA8 Fourier transform spectrometer using films on sapphire
substrates. For variable-temperature data the film was mounted in a
flow-through cryostat with polyethylene windows. The multilayer
system, sapphire-GdN-GaN, shows complex interference. Fringes with a
periodicity of about 6~cm$^{-1}$ ($\sim$0.1~meV) associated with the
substrate have been simply smoothed, leaving weak interference
associated with the film and capping layers. Despite the weak fringes
the band edges are strikingly clear, permitting an unambiguous
determination of the optical gap and its red shift in the
ferromagnetic phase.

\section{Results}
Figure~\ref{fig1} shows the frequency-dependent transmission at
temperatures of 6, 75 and 300~K. Turning first to the 300~K data it
can be seen that there is a clear onset of absorption just above 1
eV. There is no evidence of sub-gap absorption at lower energies, with
the transmission flat except for interference fringes that modulate
the transmission by 3\%. The gap of $1.31\pm0.03$~eV is estimated as
the intersection of the frequency-independent transmission below the
edge and the extrapolation from the inflection point centered on the
edge. At 75~K (immediately above T$_C$) the transmission is little
changed from 300~K, showing merely a 1-2\% fall in the low-frequency
transmission and a changing interference pattern associated with
contraction and with the temperature dependence of the optical
constants in the film and substrate. Most importantly there is only a
very weakly shifted inter-band edge as compared to the results at 300
K. In contrast the ferromagnetic phase, at 6~K, shows a very
significant change in both the absorption edge and the sub-gap
transmission. A very clear tail extends to a new edge at lower energy,
which can be seen much more clearly in a plot of the difference
between the transmissions at 75 and at 6~K, in the inset of
fig.~\ref{fig1}. Using again the intersection of extrapolated data
above and below the edge we find a gap of $0.90\pm0.03$~eV in the
ferromagnetic state. The lower-energy edge enhances the refractive
index below the gap, which then increases the reflectivity and is in
turn responsible for the reduced sub-gap transmission.

\begin{figure}
\centering{
 \includegraphics[width=8cm]{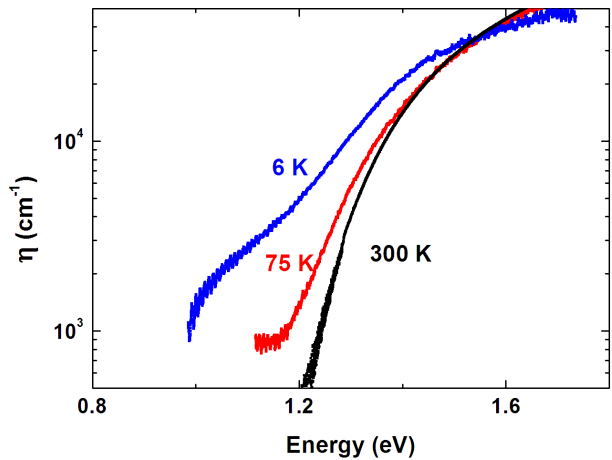}
} 
\caption{(color online) Absorption coefficient ($\eta$) implied by the data of
  fig. \ref{fig1} in paramagnetic and ferromagnetic phases.}
\label{fig3}
\end{figure}

In order to estimate the energy-dependent absorption coefficient,
$\eta$, across the edge note that, with the neglect of interference
enhancement within the GdN film, the transmission coefficient $T$ is
proportional to $(1-R)e^{-\eta d}$, where $d$ is the thickness and $R$
the reflectivity of the multi-layer. Figure \ref{fig3} shows the
absorption coefficient estimated using this equation, replacing the
factor $(1-R)$ by the transmission in the sub-gap region.  It is
significant that the absorption coefficient rises steeply to over
$10^4$~cm$^{-1}$. Such strong absorption is typical of inter-band
direct transitions, and is orders of magnitude larger than is seen at
indirect transitions.

\begin{figure}
\centering{
 \includegraphics[width=8cm]{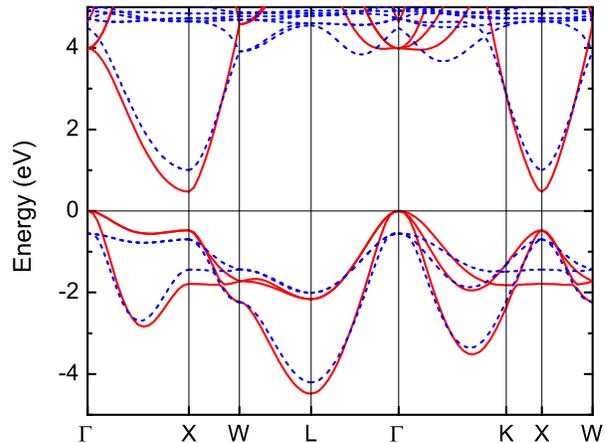}
} 
\caption{(color online) The band structure of GdN calculated with the
  value of $U_d = 8.0$~eV determined from the paramagnetic and
  ferromagnetic gaps reported in this paper. Solid lines represent
  majority spin bands, dashed lines minority.}
\label{fig4}
\end{figure}

\section{Discussion}
As discussed above there exist historical data suggesting that the optical gap in GdN is 0.98 eV.\cite{busch1969poc2} It is interesting that the same paper reported what
was regarded as anomalous optical data on one sample of GdN,
apparently with a lower-than-typical N deficiency and thus weaker free
carrier absorption, in which the diffuse absorption showed an onset
close to that reported here. The paper also made the claim that there was
no shift of the edge in the ferromagnetic phase, but that the sub-gap
absorption increased significantly. Although no low temperature data
were presented, that description suggests a behaviour similar to the present report, except that the data were apparently not extended to the red-shifted edge. 

More recently there has been a report of band shifts in GdN observed in x-ray magnetic circular dichroism data.\cite{leuenberger2006xrm} The conduction-band density-of-states (CBDOS) at the Gd L$_2$ edge was seen to red-shift by a few hundred meV in the ferromagnetic state. Those data measure the CB maximum relative to the Gd $2p$ core level, so they are not directly comparable to the present inter-band data. However, it is interesting that the red shift of the Gd 2p-to-CBDOS is of very similar magnitude to the inter-band edge red shift reported here. 

The very clear optical gaps that we report relate directly to the LSDA+\emph{U} computed
band structure of Larson \textit{et al.},\cite{larson2007esr} in which the same
red shift of 0.4~eV is predicted between the para- and ferro-magnetic
phases. However, the absolute magnitudes of the gaps are in
disagreement with our data. The experimental optical gap of Busch
\textit{et al}. \cite{busch1969poc2} was used in that calculation as a means to
fine tune the gap. This is done by choosing a $U_d$ parameter that is
applied to the empty Gd 5\emph{d} levels, shifting them up. The gap,
0.98~eV in the paramagnetic state, was assumed to correspond to the
average of the majority- and minority-spin gaps in the ground state
for which the calculation was made. From this they fix $U_{d}$ at
6.4~eV. Using the same method and software we have recalculated the
GdN band structure to fit the new data. In a remarkable agreement
between theory and experiment we find that a choice of $U_d$ = 8.0~eV
leads to a majority-spin gap of 0.91~eV and an average gap of 1.30~eV,
which correspond to our measurements within their
uncertainties. Figure \ref{fig4} shows the band structure calculated
with that parameter; note that GdN is an indirect-gap semiconductor
with the direct gap at X and an indirect ($\Gamma - X$) gap of 0.43~eV
(0.98~eV) in the ferromagnetic (paramagnetic) state.

\begin{figure}
\centering{
 \includegraphics[height=8cm]{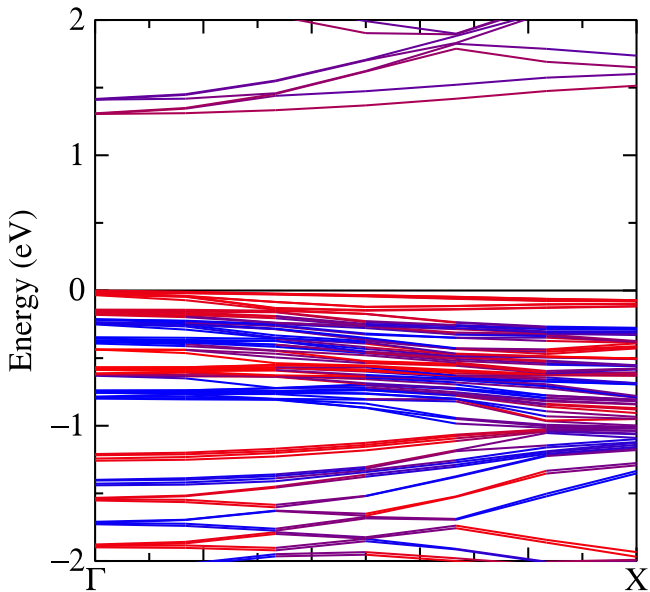}
} 
\caption{(color online) Energy bands of GdN in a $2\times2\times1$ supercell of
the cubic fcc cell (containing 32 atoms) from $\Gamma=(0,0,0)$
to $X=(\frac{\pi}{2a},0,0)$ with randomly chosen spin directions on the Gd
sites. The bands are colored according to spin-content, with red for
majority and blue for minority spin.}
\label{fig5}
\end{figure}

It is important to explore whether the representation of the gap in the paramagnetic state as the average spin-resolved ferromagnetic gap is realistic. To test the idea we have carried out calculations of both ferromagnetically-aligned and non-collinear random-aligned spins in a $2\times2\times1$ supercell of the conventional cubic fcc cell, containing 32 atoms. 
This calculation was done within the atomic sphere approximation and shifts were simply added to the Gd-$4f$ and Gd-$5d$ bands mimicking the full-fledged LSDA+\emph{U} ferromagnetic state as closely as possible. 
With this approach the average gap of spin up (1.12~eV) and spin down (1.58~eV) gaps came out to be 1.35~eV in the ferromagnetically-aligned case.
The simple shift approach compared to the LSDA+\emph{U} has as a
side-effect that the valence band maximum (VBM)  at $X$ (of rocksalt)
equals the VBM at $\Gamma$, making the direct and indirect gap equal.
However, the calculation clearly illustrates the relation between
the gap of a disordered spin arrangement and the average of spin up and
spin down gaps of the ferromagnetic system as a proof of principle.
Using the same shifts in the supercell with non-collinear spins, the gap was found to be 1.30~eV, as can be seen in fig.~\ref{fig5}, which is indeed close to but slightly less than the average of the ferromagnetic gaps. 
The reason for the small discrepancy is that the calculation was performed for only one representative random sample and the size of the cell is still relatively small to represent a random configuration. In fact, it had a net residual magnetic moment of 1.44 $\mu_B$/Gd in the cell. So, one could say it is still 20~\% magnetized instead of completely demagnetized. Thus it has only 80~\% of the upward shift in gap from the ferromagnetic case, i.e. 1.30~eV. The bands in fig.~\ref{fig5} are color coded according to their spin content. Red means 100~\% majority spin, blue means 100~\% minority spin and bands with mixed spin character have the appropriate mix of blue and red. 
One may see that the valence band maximum still has 100~\% majority spin character but the conduction band minimum has a mixed character, the lower one being slightly more majority and the next one being slightly more minority spin in character.  This direct way of simulating the paramagnetic system would require larger cells to be fully converged. 
Nevertheless, it provides a direct test that the gap is indeed larger for a system with randomly oriented spins compared to the gap for the ferromagnetic system and provides additional support to our procedure of estimating the gap of the paramagnetic system as the average of majority and minority spin gaps in the ferromagnetic state.

\section{Conclusion}
We have reported an unambiguous signature of the direct
optical gap in a rare-earth nitride. The absorption edge is seen as a
rapid loss of transmitted intensity as the photon energy rises through
the edge, and is accompanied by an energy-independent transmission
indicating no significant absorption below the edge. The direct gap is
seen at 1.3~eV in the paramagnetic phase at 300~K, and on entering the
ferromagnetic phase it falls to 0.9~eV. The red shift of 0.4~eV is in
good agreement with recent LSDA+\emph{U} calculations and a slight
fine tuning of the empirical $U_d$ parameter used in those
calculations allows us to reproduce the absolute values of the gaps in
the band structure calculation. 
The procedure of estimating the gap of the paramagnetic state as the
average of majority and minority spin gaps in the ferromagnetic state
was justified by a separate calculation for a system with randomly
oriented spins.
It is significant that earlier
predictions concerning the conducting character of the entire range of
RE nitrides were based on the value of $U_d$ in GdN, adjusted so as to
reproduce the earlier incorrect optical gap of 0.98~eV in the
paramagnetic state. Similar fine-tuning readjustments of band
structures will be required also for other RE nitrides.

\begin{acknowledgments}
The MacDiarmid Institute for Advanced Materials and
Nanotechnology is supported by a grant from the New Zealand Tertiary
Education Commission under the Centre of Research Excellence Fund. The
work at CWRU was supported by the Army Research Office under grant
number W911NF-06-1-0476.
\end{acknowledgments}

\end{document}